# Computing in the Life Sciences: From Early Algorithms to Modern AI

## Authors

- **Samuel A. Donkor**
  · [samadon1](#)
  In Vivo Group

- **Matthew E. Walsh**
  [0000-0003-1514-7761](#) · [mwalsh52](#)
  U.S. National Security Commission on Emerging Biotechnology; Department of Environmental Health and Engineering, Johns Hopkins Bloomberg School of Public Health

- **Alexander J. Titus** ✉
  [0000-0002-0145-9564](#) · [alexandertitus](#)
  In Vivo Group; U.S. National Security Commission on Emerging Biotechnology; Information Sciences Institute & Iovine and Young Academy, University of Southern California

✉ — Correspondence possible via [GitHub Issues](#) or email to Alexander J. Titus <publications@theinvivogroup.com>.

## Abstract

Computing in the life sciences has undergone a transformative evolution, from early computational models in the 1950s to the applications of artificial intelligence (AI) and machine learning (ML) seen today. This paper highlights key milestones and technological advancements through the historical development of computing in the life sciences. The discussion includes the inception of computational models for biological processes, the advent of bioinformatics tools, and the integration of AI/ML in modern life sciences research. Attention is given to AI-enabled tools used in the life sciences, such as scientific large language models and bio-AI tools, examining their capabilities, limitations, and impact to biological risk. This paper seeks to clarify and establish essential terminology and concepts to ensure informed decision-making and effective communication across disciplines.

*The views and opinions expressed within this manuscript are those of the authors and do not necessarily reflect the views and opinions of any organization the authors are affiliated with.*

# Executive Summary

## Computing in the Life Sciences: From Early Algorithms to Modern AI

The integration of computing technologies into the life sciences has revolutionized the field, enabling unprecedented advancements in biological research and applications. This manuscript traces the historical milestones and technological advancements that have shaped this transformative journey.

The early days of computing in the life sciences saw the use of primitive computers for population genetics calculations and biological modeling in the 1950s. This period marked the rise of computational biology, with computers becoming indispensable for protein crystallography and the determination of three-dimensional protein structures.

The 1960s and 1970s witnessed significant developments, including the shift from protein to DNA analysis, driven by the advent of DNA sequencing methods. Dynamic programming algorithms for sequence alignment and pioneering methods for inferring phylogenetic trees from DNA sequences emerged during this time.

The 1980s and 1990s were pivotal, characterized by parallel advancements in molecular biology and computing. Gene targeting techniques, the polymerase chain reaction (PCR), and the emergence of bioinformatics software suites propelled the field forward. The completion of the Haemophilus influenzae genome in the mid-1990s ushered in the genomic era, culminating in the publication of the human genome at the turn of the century.

The last two decades have seen the integration of artificial intelligence (AI) and machine learning (ML) into the life sciences, revolutionizing data analysis, drug discovery, and personalized medicine. AI models, from early expert systems to modern deep learning architectures, have enhanced our ability to predict protein structures, analyze genomic data, and design novel biological entities.

The manuscript delves into the various categories of AI-enabled tools used in the life sciences, focusing on large language models (LLMs) and biological design tools (BDTs). LLMs, such as GPT and BERT, have been adapted for the life sciences domain, giving rise to specialized models like scientific LLMs (Sci-LLMs), protein LLMs (Prot-LLMs), and genomic LLMs (Gene-LLMs). These models excel at tasks such as processing scientific literature, predicting protein structures and functions, and analyzing genomic data.

BDTs, on the other hand, aid in the design of proteins, viral vectors, and other biological agents. Protein structure prediction tools, like AlphaFold and RoseTTAFold, have revolutionized the field by drastically reducing the time required to determine protein structures. Other subcategories of BDTs include protein sequence design tools, small molecule design tools, vaccine design tools, and genetic modification tools, each serving specific purposes in biological research and applications.

The manuscript also highlights the importance of benchmarking and evaluating AI models in the life sciences. Bloom's taxonomy and frameworks like SciEval and KnowEval are used to assess the capabilities of LLMs across different cognitive levels and scientific knowledge domains. Specific benchmarks for Sci-LLMs, Prot-LLMs, Gene-LLMs, and multimodal Sci-LLMs are discussed, emphasizing the need for rigorous evaluation to ensure the reliability and effectiveness of these tools.

While the integration of AI in the life sciences has enabled rapid progress, it also presents potential risks and limitations. Inaccurate outputs from AI models, stemming from biased or incomplete training data, can misguide researchers and waste valuable resources. The potential misuse of AI in creating harmful biological agents raises significant biosecurity concerns. Ethical considerations, such as data privacy, informed consent, and algorithmic bias, must be addressed to ensure responsible and beneficial use of AI in the life sciences.

Looking ahead, the manuscript underscores the need for more comprehensive benchmarks that assess AI models' performance in real-world applications and their ability to adapt to evolving scientific knowledge. Techniques like red teaming, blue teaming, and violet teaming are proposed to build resilient AI systems that minimize harm and maximize benefit. The integration of Machine

Learning Security Operations (MLSecOps) is also highlighted as a crucial step in ensuring the safety and security of AI models in the life sciences.

In conclusion, the integration of computing technologies into the life sciences has transformed the field, enabling unprecedented advancements in biological research and applications. From the early days of computational modeling to the sophisticated AI-driven tools of today, this journey has been marked by historical milestones and technological breakthroughs. As we move forward, harnessing the power of AI, cloud computing, and other emerging technologies will continue to drive innovation, offering new solutions to complex biological problems and improving human health. However, navigating the challenges and ethical considerations associated with AI in the life sciences will be crucial to ensure its responsible and beneficial use.

# Introduction

Computing technologies have become indispensable to life scientists, changing how research is conducted and expanding the scope of scientific discovery. The history of computing in the life sciences is marked by significant milestones that have advanced research, including early algorithmic approaches to the application of artificial intelligence (AI) and machine learning (ML). Early uses of computers in the 1950s for population genetics calculations and the pioneering work of Alan Turing in biological morphogenesis set the stage for subsequent developments. Over the following decades, computational biology evolved from basic protein structure analysis to complex genomic studies, driven by advancements in DNA sequencing and computing.

Today, the terms AI, ML, deep learning, and large language models (LLMs) are often used interchangeably in the life sciences. Although these terms are related, they each have distinct meanings (Figure 2). AI broadly refers to machines designed to mimic human intelligence. ML is a subset of AI focused on algorithms that improve through experience. Deep learning is a subset of ML involving neural networks with many layers that can learn from vast amounts of data, and LLMs, such as GPT (Generative Pre-trained Transformer) models like ChatGPT, are a specific type of deep learning that excel in understanding and generating human-like text. By processing and analyzing biological data at unprecedented scale and speeds, these technologies have advanced fields such as bioinformatics, structural biology, and genomics. Understanding distinctions among AI-related nomenclature is crucial as technology development accelerates. Decisions about funding, regulation, new product development, and the implementation of new technologies rely on an accurate understanding of what these technologies can and cannot do. A nuanced understanding of the capabilities and limitations of AI, ML, LLMs and other computational tools can help to correctly estimate their potential and effectively utilize valuable resources.

This paper provides an overview of historical context, current applications, and future directions of computing in the life sciences. By explaining key terms, concepts, and timelines, we aim to bridge the knowledge gap between practitioners and stakeholders, fostering an environment for progress that supports scientific innovation and public benefit outcomes.

# Computers, Algorithms and the Internet

## 1950s and 1960s: Early computers and algorithms

Computers were used in the early 1950s for population genetics calculations [1]. The inception of computational modeling in biology coincides with the origins of computer science itself. British mathematician and logician Alan Turing, often referred to as "the father of computing", used primitive computers to implement a model of biological morphogenesis (the emergence of pattern and shape in living organisms) in 1952 [2]. At about the same time, a computer called MANIAC was used for measuring speculative genetic codes; it was originally built for weaponry research at the Los Alamos National Laboratory in New Mexico [3].

Computers were used for the study of protein structure by the 1960s, and other increasingly diverse analyses. These developments marked the rise of the computational biology field, stemming from research focused on protein crystallography, in which scientists found computers indispensable for carrying out laborious Fourier analyses to determine the three-dimensional structure of proteins [4,5].

In addition to advances in determination of protein structures through crystallography, the first sequence of protein, insulin, was published [6,7]. More efficient protein sequencing methods, such as the Edman degradation technique [8], enabled sequencing 15 different proteins over a decade [9]. COMPROTEIN, one of the first bioinformatics softwares developed in the early 1960s, was designed to overcome the limitations of Edman sequencing [10]. In an effort to simplify the handling of protein sequence data for the COMPROTEIN software, a one-letter amino acid code was developed [11]. This one-letter code was first used in the Atlas of Protein Sequence and Structure [12], the first biological sequence database, laying the groundwork for paleogenetic studies.

Development of methods to compare protein sequences followed. The Needleman-Wunsch algorithm [13], the first dynamic programming algorithm developed for pairwise protein sequence alignments, was introduced in the 1970s. Multiple sequence alignment (MSA) algorithms followed in the early 1980s. Progressive sequence alignment was introduced by Feng and Doolittle in 1987 [14]. The MSA software CLUSTAL, a simplification of the Feng-Doolittle algorithm [15] was developed in 1988. It is still used and maintained to this day [16].

## 1970s: From protein to DNA analysis

The deciphering of all 64 triplet codons of the genetic code in 196817 fueled a desire to efficiently determine the sequence of DNA that existed into the 1970s. This desire led to the development of cost-efficient DNA sequencing methods, such as the Maxam-Gilbert and Sanger sequencing techniques in the mid-1970s [6,7,17]. With this new ability to generate DNA sequence data, a paradigm shift from protein analysis to DNA analysis occurred in the late 1970s. Concurrently, concerns over recombinant DNA research led to safety protocols established during the 1975 Asilomar conference [18].

New DNA sequencing techniques resulted in significantly more data to be analyzed, a task at which computation could help. The first software dedicated to analyzing Sanger sequencing reads was published in 1979 [19]. DNA sequences began to be utilized in phylogenetic inference with pioneering methods like maximum likelihood for inferring phylogenetic trees from DNA sequences [20]. Several bioinformatics tools and statistical methods were developed following this work. The adoption of Bayesian statistics in molecular phylogeny in the 1990s was inspired by this [21] and is still commonly used in biology today [22]. Yet, numerous computational limitations needed to be overcome during the latter half of the 1970s to expand the utilization of computing in the life sciences, especially in DNA analysis. The subsequent decade proved instrumental in addressing these challenges.

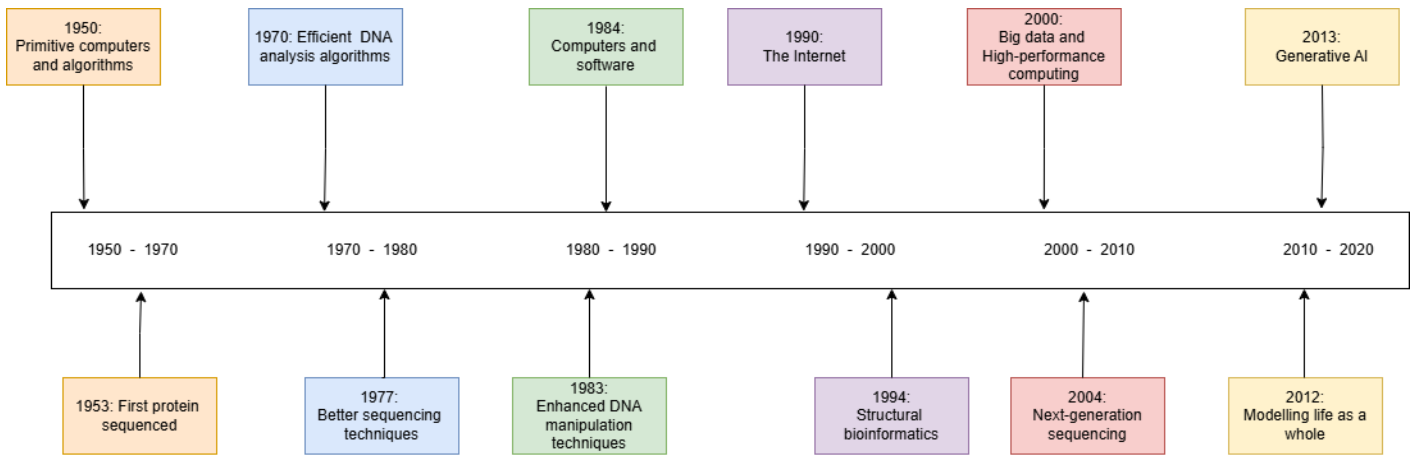

**Figure 1**: The history of parallel advancements in computing and the life sciences: A timeline of major milestones.

## 1980s: Simultaneous advances in computing and biology

Parallel advancements in biology and computing propelled bioinformatics forward during the 1980s and 1990s. Molecular techniques like gene targeting and amplification, using enzymes like restriction endonucleases and DNA ligases, laid the groundwork for genetic engineering [18]. The polymerase chain reaction (PCR) transformed gene amplification, while innovations like Taq polymerase and thermal cyclers optimized the process [23].

Computing accessibility surged with microcomputers like the Commodore PET, Apple II, and Tandy TRS-80, along with bioinformatics software like the GCG software suite [24] and DNASTAR [25], another sequence manipulation suite capable of assembling and analyzing Sanger sequencing data. Other sequence manipulation suites were developed to run on CP/M, Apple II, and Macintosh computers [26] in the years 1984 and 1985. Free code copies of this software were offered on demand by some developers. This propelled an upcoming software-sharing movement in the programming world [27,28].

The free software movement, led by the GNU project, promoted open-source bioinformatics tools. Major sequence databases (EMBL, GenBank, DDBJ) standardized data formatting and enabled global sharing. Bioinformatics journals, like CABIOS, which is now known as Bioinformatics (Oxford, England) accentuated computational methods' importance. Desktop workstations with Unix-like systems and scripting languages aided bioinformatics analyses, and scripting languages simplified tool development.

## 1990s: The genomics era and web-based bioinformatics

The genomics era began in the mid-1990s with the complete sequencing of the Haemophilus influenzae genome [29], initiating genome-scale analyses. This milestone was followed by the publication of the human genome at the beginning of the 21st century, which served as the definitive catalyst for the genomic era [30]. This transformative event spurred the design and development of several specialized Perl-based software to assemble whole-genome sequencing reads: PHRAP [31], Celera Assembler [32] among others.

Tim Berners-Lee's pioneering work at CERN in the early 1990s resulted in the World Wide Web, transforming global communication and ushering in an era of unprecedented access to information. With the advent of the internet, researchers gained a powerful platform to share and access vast amounts of biological data efficiently. This facilitated collaborative efforts in biology and genomics, leading to the establishment of foundational databases such as the EMBL Nucleotide Sequence Data

Library [33] and the GenBank database became the responsibility of the NCBI [34] in 1992. Also, the famous NCBI website came online in 1994, featuring the efficient pairwise alignment tool BLAST [35]. After that, the world saw the birth of major databases we still rely on today: Genomes (1995), PubMed (1997), and Human Genome (1999) [36,37,38].

The proliferation of web-based resources transformed access to bioinformatics tools, democratizing their availability and usability for researchers worldwide. Through the development of web platforms, bioinformatics tools became more user-friendly and accessible. This shift enabled researchers to interact with sophisticated analytical tools without needing extensive computational expertise or access to specialized hardware. Consequently, the widespread adoption of web-based bioinformatics resources facilitated broader participation in genomic and molecular research, accelerating scientific discovery and collaboration on a global scale. Graphical web servers emerged as a convenient alternative to traditional UNIX-based systems, simplifying data analysis without the need for complex installations. The continued relevance of servers for scientific purposes is exemplified by the AlphaFold Server which uses the latest AlphaFold 3 model [39], released in 2024, to provide highly accurate biomolecular structure predictions in a unified platform.

The internet facilitated the dissemination of scientific research through online publications, challenging traditional print-based methods. Early initiatives like BLEND [40] paved the way for internet-based scientific publishing by shedding insights into the potentials and obstacles associated with using the internet for scientific publications. This study paved the way for leveraging the Internet for both data set storage and dissemination, leading up to the establishment of preprint servers like arXiv (est. 1991) [41] and bioRxiv (est. 2013) [42] which changed the way scientific findings are shared and accessed. These platforms democratized access to scientific knowledge by enabling researchers to share their work rapidly and openly, facilitating interdisciplinary collaborations and the cross-pollination of ideas.

The experimental determination of the first three-dimensional structure of a protein, specifically, myoglobin, occurred in 1958 via X-ray diffraction [4]. However, earlier groundwork by Pauling and Corey with the publication of two articles in 1951 that reported the prediction of α-helices and β-sheets [43] laid the foundation for predicting protein structures. Similar to advances in other biological sciences, the utilization of computers has made it feasible to conduct calculations aimed at predicting the secondary and tertiary structure of proteins, with varying levels of confidence. This capability has been notably enhanced by the development of fold recognition algorithms, also known as threading algorithms [44,45]. However, proteins are dynamic entities, requiring advanced biophysical models to describe their interactions and movements accurately. Force fields have been formulated to describe the interactions among atoms, enabling the introduction of tools for modeling the molecular dynamics of proteins during the 1990s [46]. Used to study the behavior and interactions of atoms and molecules over time, molecular dynamics simulations calculate the positions and velocities of atoms based on physical principles. Despite the theoretical advancements and availability of tools, executing molecular dynamics simulations remained challenging in practice due to the substantial computational resources they demanded.

Graphical processing Units (GPUs) have made molecular dynamics more accessible [47], with applications extending to other bioinformatics fields requiring intensive computation. However, the internet's role in data dissemination, coupled with increasing computational power, has led to the proliferation of 'Big Data' in bioinformatics.

## 2000s: High-throughput sequencing and big data

Second-generation sequencing technologies democratized high-throughput bioinformatics. For example '454' pyrosequencing, a high-throughput DNA sequencing technique played a significant role in advancing genomics research by enabling rapid and cost-effective sequencing of DNA samples, particularly for applications such as whole-genome sequencing [48], but computational challenges arose with increased data volumes. Decreasing sequencing costs resulted in more data being generated, emphasizing data organization and accessibility. Specialized repositories and standardization efforts were needed to ensure data interoperability. High-performance computing adaptation became vital to address the increased amounts of data within bioinformatics projects. The surge in bioinformatics projects, accompanied by a vast influx of data, prompted adjustments from funding bodies to accommodate the demand for high-performance computing resources and collaborative initiatives.

While basic computer setups suffice for some projects, others demand complex infrastructures and substantial expertise. Government-sponsored entities like Compute Canada, New York State's High-Performance Computing Program, The European Technology Platform for High-Performance Computing, and National Center for High-Performance Computing served researchers' computational needs. Companies like Amazon, Microsoft, and Google, among many others, offer bioinformatics and life sciences services, emphasizing the field's importance.

## Table 1. Organizations providing High-Performance Computing Resources for Bioinformatics and Life Sciences

| Organization | Computing Resources |
|---|---|
| **Compute Canada** | Provides high-performance computing resources and support services to researchers and innovators across Canada. They offer supercomputers, cloud platforms, data storage, and training programs to advance scientific research and innovation in various fields. |
| **New York State's High-Performance Computing Program** | Provides researchers, businesses, and educational institutions with access to high-performance computing (HPC) resources and expertise to support their computational research and development efforts. |
| **The European Technology Platform for High-Performance Computing** | Fosters collaboration among industry, research, and academic stakeholders to advance high-performance computing (HPC) technology in Europe. |
| **National Center for High-Performance Computing** | Facility for high-performance computing (HPC) resources including large-scale computational science and engineering, cluster and grid computing, middleware development, visualization and virtual reality, data storage, networking, and HPC-related training. |
| **National Center for Supercomputing Applications** | Offers high-performance computing resources such as the Blue Waters supercomputer, provides advanced data storage solutions, data analysis, and visualization tools, and supports interdisciplinary research in fields such as astrophysics, climate modeling, and genomics. |
| **Oak Ridge Leadership Computing Facility** | Provides supercomputing resources, such as the Summit supercomputer, for scientific research, offers support services including software development, data storage, and visualization, and facilitates research in various fields including climate science, biology, and materials science. |
| **Swiss National Supercomputing Centre** | Provides high-performance computing systems including the Piz Daint supercomputer, offers cloud computing services, data management, and user support, and facilitates scientific research in areas such as climate modeling, physics, and life sciences. |

| Organization | Computing Resources |
|---|---|
| **Barcelona Supercomputing Center** | Provides access to MareNostrum, one of the most powerful supercomputers in Europe, offers resources for high-performance computing, data storage, and computational sciences, and supports research in fields including bioinformatics, computational biology, and engineering. |
| **Japan's RIKEN Center for Computational Science** | Houses the Fugaku supercomputer, one of the world's fastest supercomputers, provides resources for computational science, data processing, and artificial intelligence, and supports research in fields such as life sciences, materials science, and disaster prevention. |
| **National Supercomputing Centre Singapore** | Provides high-performance computing resources and support services, offers data storage, cloud computing, and software development services, and supports research in fields including bioinformatics, environmental modeling, and smart cities. |

Community computing platforms democratized participation and expanded bioinformatics research's reach. Platforms like BOINC enabled broad participation in bioinformatics. Experts can submit computing tasks to BOINC, while non-experts and science enthusiasts can volunteer their computer resources to process these tasks. Several life sciences projects are available through BOINC, including protein-ligand docking, malaria simulations, and protein folding [49].

## 2010+: The present and future

The integration of computers into biology has ushered in a new era of research possibilities, allowing for increasingly complex studies. While before, the focus was on individual genes or proteins, advancements today enable the analysis of entire genomes or proteomes [50]. This shift toward a holistic approach in biology is evident in disciplines like genomics, proteomics, and glycomics, which have limited interconnection between them.

The next leap at the intersection of computing and the life sciences lies in modeling entire living organisms and their environments simultaneously, integrating all molecular categories. This has already been achieved in a whole cell model of Mycoplasma genitalium, in which all its genes, products and their known metabolic interactions have been reconstructed [51]. Driven by advancements in measurement techniques, improved computational performance and artificial intelligence (AI) techniques, whole-cell modeling is increasingly becoming realistic and feasible. In contrast to traditional bottom-up approaches relying on molecular interaction networks, a predictive model has been developed for genome-wide phenotypes of budding yeast using deep learning [52]. The main applications of whole-cell modeling have been in producing useful substances and discovering drugs, such as antimicrobials [53,54,55,56] since whole-cell modeling was first directed toward unicellular organisms. Meanwhile, models of cultured human cells have also been developed, which have found applications in cell differentiation and medical research [57]. The possibility of modeling entire multicellular organisms may not be far off, considering the rapid pace of technological and computational advancements like artificial intelligence (AI) .

# Artificial Intelligence (AI)

Artificial intelligence (AI) refers to a set of tools, techniques and paradigms that enable computers to mimic human behavior and either replicate the decision-making process typically performed by humans or exceed human performance in solving complex tasks independently or with minimal human intervention [58]. AI is concerned with a variety of central problems, including knowledge representation, reasoning, learning, planning, perception, and communication. It also refers to a variety of tools and methods, including case-based reasoning, rule-based systems, genetic algorithms, fuzzy models, and multi-agent systems [59]. Early AI research focused primarily on hard-coded statements in formal languages, which a computer can then automatically reason about based on logical inference rules. These computer systems known as expert systems, excelled in specific domains but lacked adaptability. Over time, AI has evolved to include a variety of approaches, each with its own strengths and weaknesses. For instance, expert systems are highly accurate within narrow fields but struggle with tasks outside their programmed knowledge. In contrast, machine learning algorithms can generalize from data and adapt to new situations, though they require large datasets and extensive training. Other AI techniques, such as deep learning, neural networks, and natural language processing also offer their own unique advantages and challenges.

## Expert systems

Expert systems are a type of artificial intelligence (AI) that aims to replicate the decision-making capabilities of human experts in specific domains. They are made of a knowledge base containing domain-specific facts, rules, and heuristics, and an inference engine that applies logical reasoning to this knowledge to draw conclusions or make decisions [60]. Users are typically able to input queries and receive advice or recommendations through a simplified user interface. The primary user action, which involves pointing and clicking, is known as selecting [61].

An expert system for chemical analysis was developed in 1965 by AI researcher Edward Feigenbaum and geneticist Joshua Lederberg. This system was originally known as Heuristic DENDRAL and later as DENDRAL [62]. DENDRAL was developed to analyze molecular structures, particularly those containing elements like carbon, hydrogen, and nitrogen, based on spectrographic data. It proposed molecular structures for the compounds, with accuracy comparable to that of expert chemists.

Edward Shortliffe's work on MYCIN [63] began in 1972 at Stanford University. MYCIN, an expert system, was designed to assist physicians in diagnosing and selecting therapies for patients with bacterial infections, particularly patients with meningitis. It used a rule-based system that analyzed patient symptoms and medical history to suggest appropriate antibiotic treatments. MYCIN exhibited proficiency equivalent to infectious disease doctors.

However, despite their capabilities, the paradigm faces several limitations as humans generally struggle to explicitly articulate all their tacit knowledge that is required to perform complex tasks [64], leading to challenges such as difficulty in extrapolation, handling out-of-distribution data, managing uncertainty, and addressing biases. These limitations arise because expert systems heavily rely on predefined rules and knowledge encoded by humans. Consequently, the involvement of humans in specifying these parameters is essential but can also introduce limitations due to human cognitive constraints and biases. In contrast, machine learning algorithms overcome some of these limitations by learning from data, and making them more adaptable without relying heavily on explicit human guidance.

## Machine learning and Deep learning

Machine learning (ML) is a subset of AI that focuses on the development of algorithms and statistical models that enable computers to perform tasks without being explicitly programmed to do so [65]. It involves the use of data and algorithms to imitate the way humans learn, gradually improving the system's performance on a specific task over time through iterative learning processes. Machine learning is effective for tasks such as classification, regression, and clustering, particularly when they involve high-dimensional data. These algorithms analyze data, identify patterns, and make predictions or decisions without being explicitly programmed for each task.

Based on the given problem and the available data, there are many potential model and training paradigms, three of the most prominent types of ML being: supervised learning [66], unsupervised learning [67,68], and reinforcement learning [69]. The goal of machine learning is to develop an output model that can make predictions or decisions based on input data. In supervised learning, the model is trained on a labeled dataset, where each training example is paired with an output label. A label is the desired output or result for a given piece of data. For example, in an image recognition task, labels could be the names of objects in the images (e.g., "cat," "dog," "car"). In a spam detection task, emails could be labeled as "spam" or "not spam.". The goal is to learn a mapping from inputs to outputs. Unsupervised learning involves training a model on data without labeled responses. The goal is to uncover patterns or structures within the data. In reinforcement learning, an agent learns to make decisions by interacting with an environment. The agent receives feedback in the form of rewards or penalties based on its actions and learns to maximize cumulative rewards over time.

Depending on the learning task, the field offers various classes of ML algorithms, each of them coming in multiple specifications and variants, including regression models, instance-based algorithms, decision trees, Bayesian methods, and artificial neural networks, among others.

Artificial neural networks (ANNs) span all three major types of machine learning. ANNs are inspired by biological systems and consist of interconnected processing units called neurons, with connections akin to synapses in the human brain. Signals are processed based on thresholds set by activation functions, and organized into layers for input, hidden, and output layers. Shallow machine learning encompasses simpler ANNs and other algorithms, often being more interpretable than deep neural networks. Deep neural networks, which have multiple hidden layers, perform complex calculations to automatically discover patterns in data. This ability is known as deep learning64. Deep learning excels with large, high-dimensional data like text, images, and videos, while shallow learning may outperform with low-dimensional data or limited training data. Time series, image, and text data present various application domains.

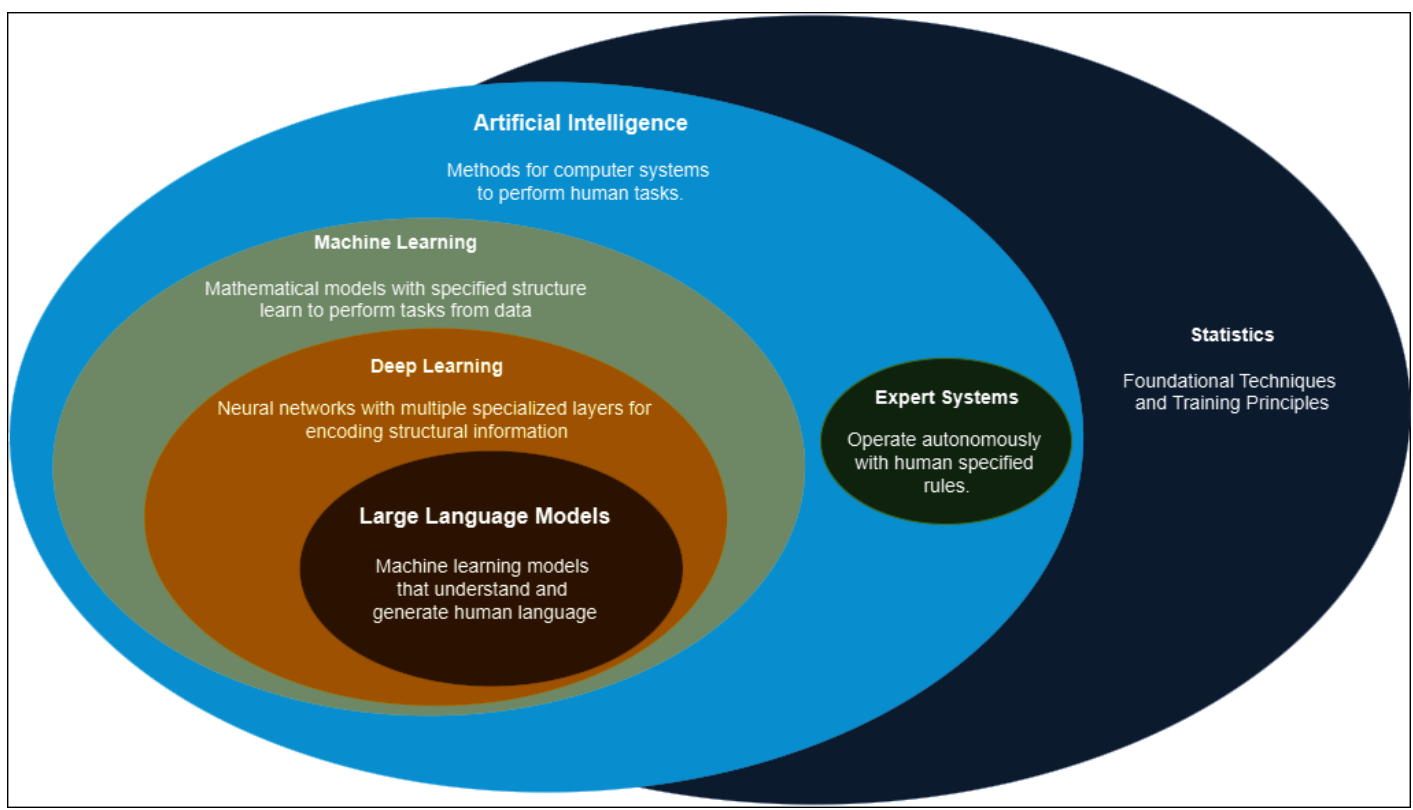

**Figure 2**: Relationship between statistics, artificial intelligence, expert systems, machine learning, deep learning and large language models.

Automated model building in machine learning involves using input data for pattern identification relevant to the learning task. Shallow machine learning relies on predefined features such as pixel values in images or word frequencies in text. For example, in image classification, shallow learning might rely on handcrafted features like color histograms or edge detectors. In contrast, deep learning can operate directly on high-dimensional raw input data, such as the raw pixel values of an image or the sequence of words in text. It automatically learns features at multiple levels of abstraction, allowing it to capture patterns in the data without the need for manual feature engineering. For instance, in image classification with deep learning, the model learns to detect edges, shapes, and textures from raw pixel data, resulting in improved accuracy [70].

Deep learning architectures often combine both aspects into end-to-end systems or extract features for use in other learning subsystems. Various deep learning architectures have emerged, including convolutional neural networks (CNNs) [71], recurrent neural networks (RNNs) [72], distributed representations [73], autoencoders [74], generative adversarial neural networks (GANs) [75], among others. CNNs excel in computer vision and speech recognition tasks, learning hierarchical features essential for image recognition. RNNs specialize in sequential data structures like time-series data and natural language processing (NLP), addressing the challenges of vanishing gradients through advanced mechanisms like long short-term memory (LSTM) networks [76]. Distributed representations, such as word embeddings, play a crucial role in NLP tasks by projecting language entities into numerical representations, preserving semantic relationships between words. Autoencoders provide dense feature representations and are applied for unsupervised feature learning, dimensionality reduction, and anomaly detection. GANs, belonging to generative models, learn probability distributions over training data to generate new data samples, using a generator-discriminator framework in a non-cooperative game setting.

## Generative AI and Transformers

Generative AI (GenAI) analyzes vast amounts of data, looking for patterns and relationships, then uses these insights to create fresh, new content that mimics the original data [77]. It does this by leveraging machine learning models, especially unsupervised and semi-supervised algorithms. There are three popular techniques for implementing Generative AI: Generative Adversarial Networks (GANs), Variational Autoencoders (VAEs), and Transformers.

Variational Autoencoders (VAEs) [78] first introduced by Diederik P. Kingma et al. in 2013 are generative models in unsupervised machine learning that generate new data similar to the input data. They consist of an encoder that compresses the input data into a lower-dimensional latent space by producing parameters for a probability distribution (mean and variance). The decoder reconstructs the data from this latent representation. The loss function, which combines reconstruction loss and regularization loss (KL Divergence), ensures the output data is both accurate and diverse. VAEs are used in applications like image generation, data imputation, anomaly detection, offering a flexible framework for generating and understanding data despite some challenges in balancing the loss components and achieving high-quality outputs [79,80,81].

In 2014, GANs [75] were proposed by researchers at the University of Montreal. GANs use two models that work in tandem: One learns to generate a target output (like an image) and the other learns to discriminate true data from the generator's output. The generator tries to fool the discriminator, and in the process learns to make more realistic outputs. The image generator StyleGAN [82] is based on these types of models.

Diffusion models [83] were introduced a year later by researchers at Stanford University and the University of California at Berkeley. By iteratively refining their output, these models learn to generate new data samples that resemble samples in a training dataset and have been used to create realistic-looking images. A diffusion model is at the heart of the text-to-image generation system Stable Diffusion [84].

Recurrent neural networks (RNNs) and their variants like long short-term memory (LSTM) networks are commonly used for sequential data processing tasks. However, these models suffer from limitations such as vanishing gradients and inefficiency in parallelization. Transformers revolutionized the field with the ability to capture long-range dependencies in sequential data efficiently and was first reported in the seminal 2017 paper, "Attention is All You Need" [85]. The introduction of transformers, with their superior performance and scalability, initiated a departure from RNNs. Transformers were used to train the large language models (LLMs) that power ChatGPT [86].

The transformer architecture consists of an encoder and a decoder, each with multiple layers of self-attention and feedforward neural networks. The self-attention mechanism enables the model to assess the significance of a piece of data, such as a word in a sentence, based on that word's relations with other words in the sentence. To preserve the ordering of the words and the meaning of the sentence, the transformer incorporates positional bias to maintain the relative positions of words within a sentence.

The transformer encoder-decoder architecture performs well at tasks like language translation. In a language translation task, the model transforms a sentence by encoding inputs from one language and then decoding outputs in another. The encoder processes the input sentence and creates a fixed-size vector representation, which the decoder then uses to generate the output sentence. The encoder-decoder employs both self-attention and cross-attention mechanisms, where self-attention is applied to the decoder's inputs, and cross-attention focuses on the encoder's output.

A prominent example of the transformer encoder-decoder architecture is Google's T5 (Text-to-Text Transfer Transformer) [87], introduced in 2019. T5 can be fine-tuned for various NLP tasks, including language translation, question answering, and summarization.Real-world applications of the transformer encoder-decoder architecture include Google Translate, which utilizes the T5 model for translating text between languages, and Facebook's M2M-10080, a multilingual machine translation model capable of translating among 100 different languages.

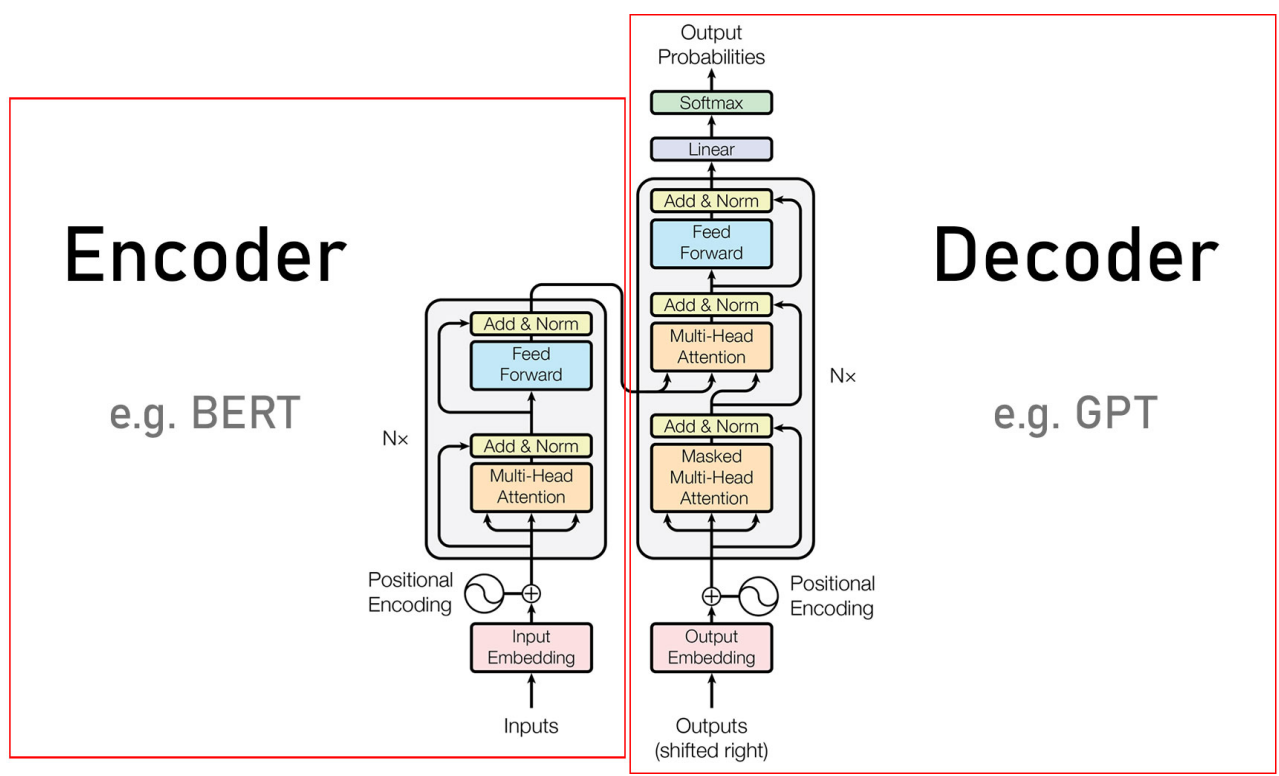

**Figure 3**: The encoder-decoder structure of the Transformer architecture. Adapted from "Attention Is All You Need" **Encoder-only models**: Ideal for tasks requiring a deep understanding of the input, such as sentence classification and named entity recognition. **Decoder-only models**: Suited for generative tasks like text generation. **Encoder-decoder models (or sequence-to-sequence models)**: Best for generative tasks that depend on an input, such as translation or summarization.

# Transformer Encoder

The transformer encoder architecture is used for tasks such as text classification, where the goal is to categorize a piece of text into predefined categories. Text classification tasks include determining the sentiment of a piece of text, determining the topic and detecting if the text is spam. The encoder processes a sequence of tokens and produces a fixed-size vector representation of the entire sequence, which is then used for classification. The most notable transformer encoder model is BERT (Bidirectional Encoder Representations from Transformers) [88], introduced by Google in 2018. BERT is pre-trained on large text datasets and can be fine-tuned for a wide range of NLP tasks.

Unlike the encoder-decoder architecture, the transformer encoder focuses solely on the input sequence without generating an output sequence and instead the output is a classification task. It uses the self-attention mechanism to identify the most relevant parts of the input for the given task. Real-world applications of the transformer encoder architecture include sentiment analysis, where models classify reviews as positive or negative, and email spam detection, where models classify emails as spam or not.

# Transformer Decoder

The transformer decoder architecture is tailored for tasks like language generation, where the model creates a sequence of words based on an input prompt or context. The decoder takes a fixed-size vector representation of the context and generates a sequence of words one at a time, with each word depending on the previously generated words. A well-known transformer decoder model is GPT-3 (Generative Pre-trained Transformer 3) [89], introduced by OpenAI in 2020. GPT-3 is a large language model capable of generating human-like text across various styles and genres. ChatGPT, which is based on the GPT-3 model, was officially launched by OpenAI in November 2020. It was a significant milestone in the development of large language models (LLMs), characterized by its ability to generate human-like text across various styles and genres. Real-world applications of the transformer decoder architecture include text generation, where models generate stories or articles based on a given prompt, and chatbots, where models create natural and engaging responses to user inputs.

# Large Language Models (LLMs)

Large language models are machine learning models that can comprehend and generate human language text. In the life sciences, LLMs such as GPT (Generative Pre-trained Transformer) and BERT, have revolutionized natural language processing, enabling researchers to extract insights from vast repositories of biomedical literature, accelerate drug discovery, and personalize patient care [90].

Large language models use transformer models and are trained using massive datasets — hence, large. This enables them to recognize, translate, predict, or generate text or other content. They are composed of multiple neural network layers – recurrent layers, feedforward layers, embedding layers, and attention layers work in tandem to process the input text and generate output content.

There are three main kinds of large language models:

- **Generic or raw language models** predict the next word based on the language in the training data. These language models perform information retrieval tasks.
- **Instruction-tuned language models** are trained to predict responses to the instructions given in the input. This allows them to perform sentiment analysis, or to generate text or code.
- **Dialog-tuned language models** are trained to have a dialog by predicting the next response. Think of chatbots or conversational AI.

Before functioning, LLMs undergo two crucial processes: training and fine-tuning. They are pre-trained on massive textual datasets from sources like Wikipedia and GitHub, comprising trillions of words to form a foundation model or a pre-trained model. This unsupervised learning stage allows the model to understand word meanings, relationships, and contextual distinctions, such as discerning whether "right" means "correct" or the opposite of "left.". To perform specific tasks, pretrained models undergo fine-tuning, which tailors them to particular activities like translation. This process optimizes task-specific performance. A related method, prompt-tuning, trains the model using few-shot or zero-shot prompting. Few-shot prompting provides examples to teach the model how to respond, while zero-shot prompting directly instructs the model on the task without examples.

LLMs serve various purposes:

- **Information retrieval**: Used by search engines like Google and Bing to produce and communicate answers conversationally.
- **Sentiment analysis**: Used to evaluate the sentiment of textual data.
- **Text generation**: Powers generative AI, such as ChatGPT, to create text based on prompts.
- **Code generation**: Similar to text generation, LLMs can generate code by recognizing patterns.
- **Chatbots and conversational AI**: Facilitate customer service interactions by interpreting and responding to customer queries.

# AI in the Life Sciences

The intersection of AI and the life sciences (AIxBio) has given rise to new capabilities where advanced computational techniques are applied to understand the complexities of biological systems and engineer novel solutions to pressing challenges in medicine and biotechnology [91]. The two primary modern AI categories used in the life sciences are large language models (LLMs) and bio-AI tools.

LLM-based chatbots like ChatGPT are designed to process human language inputs and generate output in human-like fashion. In the life sciences, ChatGPT can assist researchers by drafting and editing scientific manuscripts, generating hypotheses, summarizing datasets, and retrieving information from the scientific literature. LLM-based chatbots can also streamline literature reviews and facilitate the comprehension of complex biological concepts.

As a general-purpose LLM, ChatGPT and its equivalents are trained on a broad range of text from the internet. This results in models that function across topics and contexts. However, the generalist nature comes at the cost of precision and depth required for highly specialized tasks. For example, LLM-based chatbots can provide outputs with information with unfounded details, aiming to fill knowledge gaps. This behavior is known as "Confabulation", and it can limit the utility of the tool. Furthermore, ethical concerns related to biased outputs are often attributed to biases within the training data.

Additionally, training and using general-purpose language models can be computationally expensive, time-consuming, and resource and energy intensive. Given the cost of training general purpose LLMs and their limitations, evaluations are essential for understanding their performance. Evaluations help developers identify strengths and weaknesses of the model, and often measure generalizability of models to real-world applications. This process can also identify biased or misleading model outputs. Typically, models undergo evaluation on standardized benchmarks such as GLUE (General Language Understanding Evaluation) [92], SuperGLUE [93], HellaSwag [94], TruthfulQA [95], and MMLU (Massive Multitask Language Understanding) [96] using established metrics, as shown in Table 2.

## Table 2. Common Benchmarks for LLMs

| Benchmark | Description | Format of Task |
| --- | --- | --- |
| MMLU | MMLU (Massive Multitask Language Understanding) evaluates how well the LLM can multitask | Multiple-choice |
| TruthfulQA | Measures truthfulness of model responses | Generation, Multiple-choice |
| HellaSwag | Evaluates how well an LLM can complete a sentence | Sentence completion |
| SuperGLUE Benchmark | Compares more challenging and diverse tasks with GLUE, with comprehensive human baselines | Sentence- and sentence-pair classification (main task), coreference resolution and question answering |
| GLUE Benchmark | GLUE (General Language Understanding Evaluation) benchmark provides a standardized set of diverse NLP tasks to evaluate the effectiveness of different language models | Classification and prediction |

The behavior of LLMs can be modified through model alignment, domain-specific pre-training, and supervised fine-tuning. These methods can be used to address limitations of generic LLMs, tailor behavior to meet specific requirements, and infuse general knowledge into the LLMs. Domain-specific language models, trained or fine-tuned on specific datasets relevant to particular domains, offer more contextually accurate responses for specific domains. Evaluating domain-specific or fine-tuned models typically involves comparing their performance against a ground truth dataset if available. This process is crucial because it ensures that the model performs as expected and generates the desired outputs.

In the life sciences, specialized models can interpret complex biological data, provide detailed insights, thereby enhancing both the accuracy and reliability of the information provided. These models are known as scientific large language models (Sci-LLMs) [97].

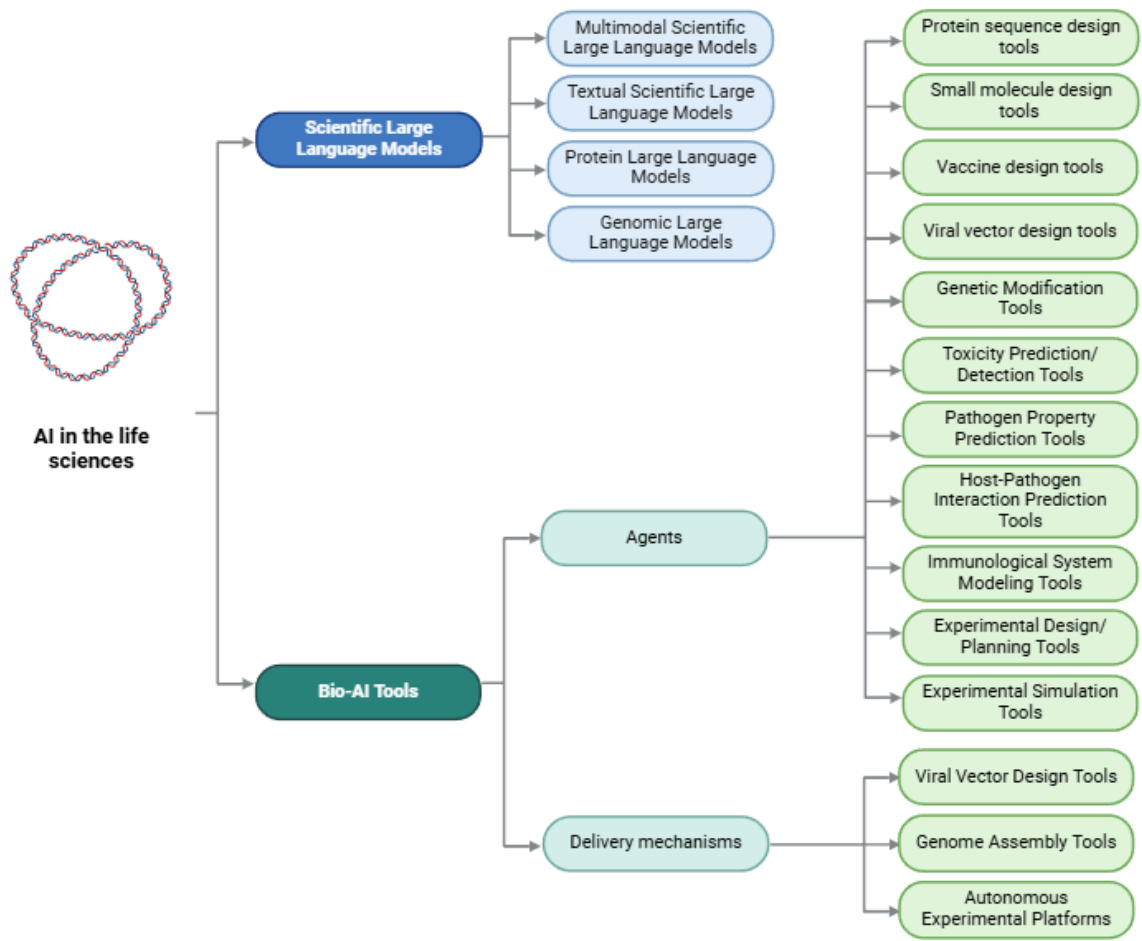

**Figure 4**: AI-enabled tools used in the biological sciences; Large Language Models (LLMs) and Bio-AI.

## Scientific Large Language Models (Sci-LLMs)

LLMs in the life sciences have been trained on natural language, molecular, protein, and genomic sequence data. These LLMs are collectively known as Scientific Large Language Models (Sci-LLMs). Sci-LLMs are specialized models designed to process and understand various types of scientific data. They extend the capabilities of general LLMs to handle domain-specific tasks in biology, chemistry, and other scientific fields. Sci-LLMs in the biological field include Textual Scientific Large Language Models (Text-Sci-LLMs), Protein Large Language Models (Pro-LLMs), and Genomic Large Language Models (Gene-LLMs) [97].

## Textual Scientific Large Language Models (Text-Sci-LLMs)

Text-Sci-LLMs are trained on vast amounts of scientific textual data, such as scientific publications. Text-Sci-LLMs excel at understanding, generating, and interacting with written human language from scientific domains. LLMs trained on vast, diverse datasets, such as BERT [88] and its variations which have been fine-tuned specifically on biological corpora with the encoder-only architecture, have demonstrated significant potential in natural language processing (NLP) tasks within biology. Models initially trained on broad corpora such as Wikipedia and textbooks and then fine-tuned on specific biological NLP tasks, show substantial improvements in various downstream tasks including biological terminology understanding, named entity recognition, text similarity, and relation extraction [98,99,100,101,102].

GPT and its variants [89,103,104], with decoder-only architectures, have become dominant in the field of biological NLP because they can generate textual information as an output. BioGPT [105], an extension of GPT-2 [104], has been extensively fine-tuned on biomedical literature, showcasing remarkable performance in biomedical relation extraction and question answering. It also generates coherent and fluent descriptions within the biomedical context. BioMedGPT-LM [106], incrementally pre-trained on LLaMA2 [107], enables a comprehensive understanding of various biological modalities and aligns them with natural language. BioGPT and BioMedGPT-LM are both specialized language models designed for biomedical applications; however, BioGPT focuses on generating and understanding biomedical literature, while BioMedGPT-LM integrates a broader range of tasks including text generation, question answering, and classification within the biomedical domain.

## Capabilities Evaluation

The evaluation of LLMs often uses Bloom's taxonomy [108,109], which includes six cognitive levels:

## Table 3. Bloom's Taxonomy

| Cognitive Level | Description | Examples of Activities/Tasks |
|---|---|---|
| Remember | Recall facts and basic concepts | List, define, identify, memorize, repeat, state |
| Understand | Explain ideas or concepts | Describe, explain, interpret, summarize, paraphrase, discuss |
| Apply | Use information or existing knowledge in new contexts | Use, demonstrate, solve, implement, execute, carry out |
| Analyze | Explore connections, causes, and relationships among ideas | Differentiate, organize, relate, compare, contrast, examine |
| Evaluate | Justify a decision or course of action based on sound analysis | Judge, critique, recommend, justify, assess, appraise |
| Create | Produce new or original work using existing information | Design, assemble, construct, develop, formulate, author |

SciEval [110] has recently introduced a framework for evaluating scientific LLMs across four dimensions: basic knowledge, knowledge application, scientific calculation, and research ability. These dimensions are based on the cognitive domains in Bloom's taxonomy. KnowEval [97] assesses the depth of knowledge LLMs can grasp, aiming for human-level comprehension. KnowEval categorizes Text-Sci-LLMs into Pre-college, College, and Post-college levels based on the complexity of scientific knowledge.

## Table 4. Categories for KnowEval

| Category | Description |
|---|---|
| Pre-college Level | This level covers fundamental concepts and principles, aligning with the Remember and Understand stages of Bloom's taxonomy and the basic knowledge dimension of SciEval. Evaluations focus on basic knowledge comprehension, using benchmarks like MMLU [96] and C-Eval [111] |

| Category | Description |
|---|---|
| College Level | At this level, knowledge becomes more specialized and abstract, requiring logical reasoning and proof. It corresponds to the Apply and Analyze stages of Bloom's taxonomy and the knowledge application and scientific calculation dimensions of SciEval. Evaluations like PubMedQA [112] and SciQ [113] focus on this advanced understanding. |
| Post-college Level | This level involves mastering current knowledge and generating innovative ideas, aligning with the Evaluate and Create stages of Bloom's taxonomy and the research ability dimension of SciEval. It requires capabilities beyond standard question-answering, including summarizing advancements and designing novel experiments. Few benchmarks, such as a subset in the SciEval dataset [110], assess these high-level capabilities. |

## Benchmarks for Text-Sci-LLMs

## Table 5. Summary of Benchmarks for Text-Sci-LLMs

| Benchmark | Description | Type |
|---|---|---|
| MMLU | Offers a detailed and challenging benchmark that tests the comprehension and problem-solving capabilities of LLMs across a wide spectrum of tasks and subjects. | Multiple choice |
| C-Eval | Consists of 13,948 multi-choice questions spanning 52 diverse disciplines and four difficulty levels. | Multiple choice |
| AGIEval | Evaluates the general abilities of foundation models in tasks pertinent to human cognition and problem-solving. | Multiple choice |
| ScienceQA | A dataset designed for question answering in the scientific domain, covering various scientific topics and requiring reasoning over structured and unstructured information. | Multiple choice / Question answering (QA) |
| SciEval | A benchmark dataset for evaluating language models in the scientific domain, covering a range of tasks related to scientific text understanding and generation. | Multiple choice / Question answering (QA) |
| Bioinfo-Bench-QA | A benchmark dataset focused on question answering in the field of bioinformatics, covering topics related to biological information processing and analysis. | Multiple choice |

| Benchmark | Description | Type |
|---|---|---|
| SciQ | A dataset designed for evaluating language models in scientific question answering tasks, covering various scientific disciplines and requiring both factual and reasoning-based answers. | Multiple choice |
| ARC | A dataset that challenges models with questions that require a mix of comprehension and reasoning skills across a wide range of topics, including science. | Multiple choice |
| BLURB | A comprehensive set of datasets and tasks designed to evaluate the performance of natural language processing (NLP) models specifically in the biomedical domain. | Multiple NLP tasks |
| PubMedQA | A dataset designed for question answering based on biomedical literature available on PubMed, aiming to evaluate models' ability to comprehend and extract information from scientific articles. | True or False |

## Protein Large Language Models (Prot-LLMs).

Protein Large Language Models (Prot-LLMs) are trained on protein-related sequence data, including amino acid sequences, protein folding patterns, and other biological information. As a result, they can accurately predict protein structures, functions, and interactions. Prot-LLMs can be categorized into three main types based on their architectures: encoder-only, decoder-only, and encoder-decoder models, each suited for various protein research applications. For instance, encoder-only models are primarily used for predicting protein functions or properties, while decoder-only models are mainly employed for protein generation tasks.

**Encoder-only models**: Encoder-only models are a specialized form of the transformer architecture, dedicated solely to understanding and encoding input sequences.The essence of an encoder-only model revolves around extracting significant context from input sequences. These models encode protein sequences into fixed-length vectors for tasks like pattern recognition and prediction. Techniques like the Pairwise Masked Language Model (PMLM) [114] and mixed-chunk attention aim to capture co-evolutionary information and reduce complexity. Non-parametric models like ProteinNPT [115] handle sparse labels and multitask learning. Some models, like ESM-GearNet [116] and LM-GVP [117], integrate 3D structure information for better performance.

**Decoder-only models**: Utilizing the GPT [89] architecture, these models, such as ProGen [118] and ProGen2 [119], are essential for controllable protein generation. They explore unseen regions of the protein space while designing proteins with nature-like properties. Similar capabilities are exemplified by models like RITA [120], PoET [121], and LM-Design [122].

**Encoder-decoder models**: Used for sequence-to-sequence tasks, these models, including ProstT5 [123] and pAbT5 [124] are adept at tasks where an input sequence is transformed into an output sequence. A common example of a sequence-to-sequence task is machine translation, where a model translates a sentence from one language to another. In the context of Prot-LLMs, sequence-to-sequence tasks could involve tasks such as translating between protein sequences and structures.

They can incorporate Multiple Sequence Alignment (MSA) modules to improve sequence generation and utilize reinforcement learning for structure-based design, as seen in Fold2Seq [125].

## Capabilities Evaluation

Prot-LLMs are evaluated in three key areas: protein structure prediction, protein function prediction, and protein sequence generation.

**Protein Structure Prediction**: Prot-LLMs can predict the 3D structure of proteins from their sequences, which aids in understanding protein function, drug design, and biomedical research. Based on the 3D structure of known proteins, prot-LLMs can predict the three-dimensional structure of proteins based on an input sequence, which includes determining the atomic coordinates and the spatial relationships between atoms. Encoder-based Prot-LLMs are used to extract sequence information from the training data and predict tertiary and quaternary structures.

**Protein Function Prediction**: Prot-LLMs can predict the biological function of proteins and their interactions with other biomolecules. These tasks can be grouped into several categories. Firstly, protein classification involves categorizing proteins based on their structure, function, or sequence similarity. Prediction of protein-protein interactions focuses on identifying and forecasting interactions crucial for various biological processes. Localization and homology detection tasks include predicting a protein's subcellular location and identifying distant relationships between protein sequences. Spectral characteristics and stability prediction involve forecasting fluorescence properties and stability under specific conditions, respectively. Furthermore, specific tasks such as $\beta$-Lactamase activity prediction, solubility prediction, and mutation effect prediction focus on understanding specific protein functions, compound solubility, and the effects of genetic mutations on protein function, respectively. These tasks collectively contribute to explaining the complex functions and behaviors of proteins in biological systems. Biological systems are inherently complex and multifaceted, often requiring the simultaneous optimization of multiple properties. Unlike single-objective optimization, which focuses on one specific goal, multi-objective optimization allows researchers to consider and balance several objectives at once. This is particularly important in protein function prediction, where factors such as stability, activity, solubility, and interaction with other molecules need to be optimized concurrently. By providing a more comprehensive optimization framework and utilizing techniques such as Pareto optimization, researchers can identify solutions that offer the best trade-offs among different objectives, rather than a single optimal solution for one objective. multi-objective optimization can enhance the practical applicability of Prot-LLMs, leading to more effective and efficient solutions in understanding and manipulating protein functions.

**Protein Sequence Generation**: Prot-LLMs can propose amino acid sequences not found in nature and with a predicted function, useful in drug design and enzyme engineering. It includes:

- De novo protein design: Proposing protein sequences with a desired property that are not based on existing proteins with some or all of the desired property. Autoregressive generative models, such as the ProGen series, are commonly utilized for tasks involving the generation of protein sequences.
- Protein sequence optimization: proposing modification to an existing protein sequence to alter (i.e., optimize) its function or characteristic in an intended manner.

## Benchmarks for Prot-LLMs

### Table 6. Summary of Benchmarks for Prot-LLMs

| Benchmark | Description |
|---|---|

| Benchmark | Description |
|---|---|
| CASP | CASP (Critical Assessment of Structure Prediction) evaluates different methods and algorithms for protein structure prediction, providing a standard assessment for progress in the field. |
| EC | EC (Enzyme Commission) dataset is used to classify enzymes based on the chemical reactions they catalyze. This system is used to evaluate the functional prediction of proteins, specifically enzymes. |
| GO | GO (Gene Ontology) provides a framework for the representation of gene and gene product attributes across species. GO terms are used to annotate proteins with their associated biological processes, cellular components, and molecular functions. |
| CATH | CATH (Class, Architecture, Topology, Homologous superfamily) is a protein structure classification database that organizes protein domains into a hierarchical structure based on their folding patterns. It is used to classify protein domains into these categories: Class, Architecture, Topology, Homologous superfamily. |
| SCOP | SCOP (Structural Classification of Proteins) classifies proteins based on their structural and evolutionary relationships. SCOP benchmarks evaluate the ability of computational methods to classify protein structures into appropriate categories: Class, Fold, Superfamily, and Family. |
| ProteinGym | ProteinGym is a benchmark suite designed for evaluating the generalization capabilities of machine learning models in protein sequence prediction tasks. It includes various datasets and metrics to assess the performance of models in predicting protein sequences and related properties under different conditions. |
| TAPE | TAPE (Task Assessing Protein Embeddings) is a benchmark suite designed to evaluate the performance of protein sequence embeddings learned by machine learning models. It includes a variety of tasks, such as secondary structure prediction, contact prediction, and remote homology detection, to assess how well these embeddings capture the underlying biological properties of proteins. |

## Genomic Large Language Models (Gene-LLMs)

Gene-LLMs, specialized in genomic data, are trained to comprehend and predict genetic and genomic aspects of biology. They analyze DNA sequences, interpret genetic variations, and aid in genetic research, like identifying disease-related genetic markers or exploring evolutionary biology. Built on the Transformer architecture, genomic LLMs effectively model nucleic acid sequence data, capturing long-range dependencies for prediction and generation tasks. Through self-supervised learning on genomic sequences, Gene-LLMs gradually grasp genome understanding. Once fine-tuned or contextually learned, they prove valuable for downstream tasks, enhancing accuracy and reducing manual intervention.

**Encoder-only models**: With an encoder-only architecture for genomics, numerous significant models utilize the Transformer encoder to process gene sequences and extract meaningful patterns. Models like SpliceBERT, DNABERT, DNABERT-2, iEnhancer-BERT [126,127,128,129], and others employ mask training mechanisms to predict and complete masked gene sequences, achieving improved performance in tasks such as promoter prediction and transcription factor binding site prediction.

For instance, MoDNA [130] adopts a BERT-like encoder with a unique stacked Generator-Discriminator training paradigm, facilitating motif-oriented learning. GENA-LM [131] introduces encoder-based foundational DNA language models capable of handling sequences up to 36,000 base pairs. The Nucleotide-Transformer model [132], pre-trained on diverse human and species genomes, enhances the prediction of molecular phenotypes from DNA sequences. EpiGePT [133] predicts genome-wide epigenomics signals, offering insights into gene regulation. Uni-RNA [134] predicts RNA structures and functions, useful in RNA research and drug development. Models like Enformer [135] and LOGO [136] address the quadratic time complexity of attention mechanisms in handling long sequences, while BioSeq-BLM [137] integrates traditional analysis methods with language models, marking advancements in pre-training and fine-tuning.

**Decoder-only models**: Decoder-only models, like GenSLMs [138] and DNAGPT [139], demonstrate generative capabilities, capturing the evolutionary dynamics of viruses and enabling species identification and regulatory factor prediction. HyenaDNA [140] stands out for its exceptional ability to efficiently handle ultra-long DNA sequences while preserving single-nucleotide resolution. This unique combination of features enables researchers to analyze and manipulate genetic data at an unprecedented level of detail. Its capability to handle long sequences while maintaining single-nucleotide resolution greatly enhances its utility in various genomic applications, representing a significant advancement in computational genomics.

**Encoder-decoder models**: Encoder-decoder models in genomics, such as ENBED [141], combine the strengths of both components to compress and encode genomic data into meaningful representations. These representations are then used by the decoder to generate sequences or make predictions, enhancing bioinformatics research capabilities.

## Capabilities Evaluation

Gene-LLMs undergo evaluation across four key domains: function prediction, structure prediction, sequence generation, and sequence variation and evolution analysis.

**Protein Function Prediction**: Traditionally, gene function prediction relied on models trained on specific sequences. With the advent of LLMs, pre-training on extensive genomic data followed by task-specific fine-tuning has enhanced accuracy and contextual understanding. Key subtasks include promoter prediction, enhancer prediction, and binding site prediction, tackled by models like DNABERT [127] and EpiGePT [133]

**Structure Prediction**: Leverages computational tools to identify and model biologically significant nucleic acid structures, aiding in the design of novel molecular architectures for nanotechnology and synthetic biology. Recent advancements include predicting RNA three-dimensional structures directly from sequences and designing sequences for predefined DNA and RNA nanostructures, demonstrating that nucleic acid structure can be both predictable and controllable. Subtasks include chromatin profile prediction and DNA/RNA-protein interaction prediction, addressed by models like HyenaDNA [140] and TFBert [142].

**Sequence Generation**: Proposing artificial sequences resembling real biological ones is crucial for bioinformatics, particularly for creating artificial human genomes serving as tools to safeguard genetic privacy and reduce costs linked with genetic sample collection [143,144]. The generated data strives to retain the utility of the source data by replicating most of its characteristics. Consequently, they could serve as viable alternatives for many genomic databases that are either not publicly available or face accessibility barriers. DNAGPT [139] excels in this task, generating artificial genomes covering regions of single nucleotide polymorphisms (SNPs).

**Sequence Variation and Evolution Analysis**: Understanding biological sequence variation and evolution is vital for uncovering the genetic basis of traits, disease, and evolutionary patterns. Models like GenSLMs [138] and GPN-MSA [145] analyze the evolutionary landscape of genomes, focusing on species-specific and whole-genome sequence alignments.

## Benchmarks for Gene-LLMs

## Table 7. Summary of Benchmarks for Gene-LLMs

| Benchmark | Description |
|---|---|
| CAGI5 Challenge Benchmark | The Critical Assessment of Genome Interpretation (CAGI) is a benchmark designed to rigorously assess computational methods in predicting a wide array of genetic and genomic outcomes. |
| Protein-RNA Interaction Prediction Benchmark (Protein-RNA) | A set of 37 machine learning (primarily deep learning) methods for in vivo RNA-binding proteins RBP–RNA interaction prediction. This benchmark systematically evaluates a subset of 11 representative methods across hundreds of CLIP-seq datasets and RBPs. |
| Nucleotide Transformer Benchmark (NT-Bench) | A comprehensive evaluation framework designed to assess the performance of genomics foundational models. This benchmark pits the Nucleotide Transformer models against other prominent genomics models, such as DNABERT, HyenaDNA (with both 1kb and 32kb context lengths), and Enformer. |

# Multimodal Scientific Large Language Models (MM-Sci-LLMs)

Multimodal scientific large language models (MM-Sci-LLMs) possess the ability to process and combine various types of scientific data, including text, molecules, and proteins, making them indispensable for interdisciplinary research requiring insights from multiple domains. An emerging research area, MM-Sci-LLMs utilize LLMs as their core to handle diverse data types effectively. These models exhibit remarkable adaptability in incorporating text, images, audio, and other forms of information, enabling comprehensive problem-solving across scientific domains, particularly in biological sciences encompassing protein, molecular, and genomic studies.

Categorized into four distinct groups based on the specific modality they focus on, MM-Sci-LLMs demonstrate specialized capabilities.

## Table 8. Summary of MM-Sci-LLMs

| Category | Description | Encoder-only models | Encoder-Decoder models | Decoder-only models |
|---|---|---|---|---|

| Category | Description | Encoder-only models | Encoder-Decoder models | Decoder-only models |
|---|---|---|---|---|
| Molecule-to-text | Leverage various techniques like multimodal embedding and cross-modal learning to associate chemical structures with textual descriptions, enhancing tasks such as cross-modal retrieval and molecular property prediction. | Text2Mol, KV-PLM, MoMu | DrugChat, MolReGPT, Text+Chem, ChatMol, GIT-Mol | MolET5, MolFM, GPT-Mol |
| Protein-to-text models | Utilize textual data for protein function prediction and multimodal representation learning, enriching protein annotation and design by integrating natural language descriptions with protein data. | ProTranslator, ProtST-ProtBert | InstructionProtein | ProteinDT, Prot2Text, ProtST-ESM-1B, ProtST-ESM-2 |
| Protein-to-molecule models | Focus on linking protein sequences with molecular information, improving drug discovery through techniques like adversarial networks and contrastive learning. | DrugCLIP | DrugGPT | ChemBERTaLM, DeepTarget |
| Comprehensive models | Integrate multiple scientific modalities to excel in diverse tasks like biological data analysis, and material prediction, leveraging advanced multimodal learning techniques to support fundamental science research. | BioTranslator | Galactica, ChatDrug, DARWIN-MDP, BioMedGPT-10B, Mol-Instructions | BioT5 |

## Capabilities Evaluation

MM-Sci-LLMs undergo evaluation focusing on three pivotal areas: cross-modal prediction, retrieval, and generation.

**Cross-Modal Prediction**: This involves using multimodal models to predict the functionality of biological entities like molecules, proteins, and genomes based on textual instructions. Models like MoleculeSTM [146] and Mol-Instructions [147] integrate molecular structures and text data for function prediction, which is crucial for bioinformatics and drug discovery.

**Cross-Modal Retrieval**: Involves retrieving information from one modality based on a query from another modality. Key models like KV-PLM [148] and ProtST-ESM-1b [149] enable retrieving molecules, proteins, or genes based on textual descriptions, aiding drug discovery and biological mechanism understanding.

**Cross-Modal Generation**: Aims to create data in one modality based on data from another. Models like Text2Mol [150] and ProteinDT [151] generate molecular information from text descriptions, while models like Prot2Text [152] and ChemBERTaLM [153] convert protein sequences into detailed text descriptions. This capability facilitates cohesive multi-modal data creation, bridging the gap between different modalities in scientific research.

## Benchmarks for MM-Sci-LLMs

## Table 9. Summary of Benchmarks for MM-Sci-LLMs

| Benchmark | Description |
| --- | --- |
| MoleculeNet | MoleculeNet is a large-scale benchmark for molecular machine learning. It curates multiple public datasets, establishes metrics for evaluation, and offers high-quality open-source implementations of multiple previously proposed molecular featurization and learning algorithms. |
| MARCEL | MARCEL (MoleculAR Conformer Ensemble Learning) provides a comprehensive platform for evaluating learning from molecular conformer ensembles. It focuses on diverse molecular conformer structures, marking a significant shift in molecular representation learning. |
| GuacaMol | GuacaMol is an evaluation framework designed for de novo molecular design. It aims to generate molecules with specific property profiles through virtual design-make-test cycles. |

Our technical exploration is primarily confined to Transformer-based languages, excluding alternative neural architectures like graph neural networks and diffusion models, despite their widespread applications in protein folding. However, the concepts discussed in biological languages can be extended to other scientific languages, such as molecular and mathematical languages.

Molecular large language models (Mol-LLMs) are specialized LLMs trained on molecular data, enabling them to understand and predict the chemical properties and behaviors of molecules. This specialized knowledge makes them invaluable tools in drug discovery, materials science, and the study of complex chemical interactions.

Encoder-only Mol-LLMs, like SMILES-BERT [154], focus on understanding and interpreting input molecules, making them ideal for tasks requiring a deep comprehension of molecular structures and properties. SMILES-BERT, for instance, leverages the BERT architecture to interpret SMILES representations of molecules.

Decoder-only Mol-LLMs, such as MolGPT [155] and SMILESGPT [156], use SMILES strings as input to navigate the vast chemical space. These models are crucial in drug discovery and materials science, enabling the synthesis of molecules with specific properties. MolGPT, which utilizes GPT for molecular generation with conditional training for property optimization, excels in molecular modeling and drug discovery by demonstrating strong control over multiple properties for accurate generation.

In encoder-decoder Mol-LLMs, encoders convert raw molecules into latent vectors, which decoders then reconstruct into functional chemical structures. Most Transformer-based encoder-decoder models use SMILES or SELFIES as inputs for the encoder, with outputs varying by task. For example, in chemical reaction prediction, the decoder generates the anticipated outcomes for reactants. The Molecular Transformer [157], a Transformer-based model for reaction prediction, effectively handles complex, long-range sequence interactions.

Biological data with graph structures can be modeled in two primary ways: molecular structure-based modeling and biological network-based modeling. In molecular structure-based modeling, atoms or valid chemical substructures are used as nodes, and bonds serve as edges to construct the molecular graph. Molecular graphs are extensively used for predicting molecular properties and designing new molecules.

In biological network-based modeling, nodes represent various entities such as genes, diseases, or RNAs, with edges indicating known associations between pairs of entities, such as miRNA–disease interactions. This creates a relational network. Graph Neural Networks (GNNs) excel at extracting information from graph structures, making them suitable for processing omics data in fields such as genomics, proteomics, RNomics, and radiomics. By applying GNNs to these omics data using the aforementioned modeling methods, a variety of tasks can be performed, including molecular property prediction, de novo molecular design, link prediction, and node classification in biological networks.

## Bio-AI Tools (BDTs)

Bio-AI tools, commonly referred to as biological design tools (BDTs) are computational tools that help design proteins, viral vectors, or other biological agents. Traditional methods molecular biology like site-directed mutagenesis (SDM) involve the deliberate alteration of specific nucleotide sequences in DNA to create desired changes in the resulting protein. This process typically requires designing and synthesizing specific DNA primers, followed by PCR amplification and cloning steps to introduce the mutated DNA into a host organism. While SDM allows for precise modifications at predetermined sites, it can be time-consuming and labor-intensive, especially when multiple iterations are required to achieve the desired outcome. Additionally, the success rate of SDM experiments can vary depending on factors such as the efficiency of DNA synthesis and the stability of the resulting mutant proteins.

Random mutagenesis, another traditional method, involves introducing random mutations throughout the genome of an organism using techniques such as chemical mutagenesis or UV irradiation. This approach generates a pool of mutants with diverse genetic variations, which are then screened to identify individuals with desired phenotypic traits. While random mutagenesis can uncover novel genetic variants and phenotypes, it lacks the precision and control offered by targeted mutagenesis techniques like SDM. A related concept that enhances the utility of random mutagenesis is directed evolution. Directed evolution is an iterative process where organisms undergo random mutations, are tested against a screening process, and the best performers are selected for subsequent rounds of mutation. This cycle of mutating, screening, and selecting can be analogized to the training process of deep learning models. In deep learning, a model makes predictions based on input data, receives feedback on the accuracy of these predictions, and then adjusts its parameters through a process known as backpropagation.

In directed evolution, the organism's genetic material is repeatedly altered and tested, much like a model's parameters are iteratively refined to improve performance. Each cycle of directed evolution involves creating genetic diversity through random mutations, screening the resultant mutants for desirable traits, and then selecting the top performers for the next round of mutations. This method has been instrumental in fields such as enzyme engineering, where it has led to the development of proteins with enhanced or novel functions. However, It is resource-intensive, requiring significant time and high-throughput screening capabilities.

In contrast, BDTs can accelerate experimentation by suggesting optimized properties of biological agents upfront, thereby potentially reducing the number of tests required to achieve desired outcomes. While the speed of individual experiments may not change, the efficiency of the overall experimentation process is enhanced, as researchers may need to conduct fewer experiments to reach the same or improved results [158]. Examples of BDTs include RFDiffusion [159], Protein MPNN [160], and protein language models like ProGen2 [119] and Ankh [161]. These models can be considered both Prot-LLMs and specific instances within the broader category of BDTs due to their training and output characteristics.

A crucial difference between LLMs and BDTs is both the training data — as LLMs are trained on natural language while BDTs are trained on biological data — and the output — LLMs typically produce outputs in natural language while BDTs produce outputs in the form of biological sequences, structures, and predictions. Although BDTs currently focus on creating sequences by optimizing for a single function, they may eventually evolve to design complex proteins and enzymes with multiple functions and properties. BDTs may eventually develop the capability to engineer whole organisms optimized for various functions and characteristics, addressing a comprehensive range of biological properties [162].

Of all the categories of AI-enabled BDTs, protein structural prediction tools have the highest relative maturity. Protein structure prediction tools, commonly referred to as 'folding tools,' contribute to the field by predicting a protein's 3D structure, including its secondary, tertiary and quaternary structures from its amino acid sequences. This prediction aids in understanding protein function and interactions. Determining the precise structure of proteins, vital for their functions, has historically posed significant challenges in experimental biology [163], often requiring years of dedicated effort. However, the landscape has shifted with the advent of AI, tailored to predict protein structures directly from their amino acid sequences.

Notably, pioneering AI systems like AlphaFold [164] and RoseTTAFold [165] have emerged, revolutionizing the field by drastically reducing structure determination times from months to mere hours. While AlphaFold provides measured structures based on experimental data and computational predictions, RoseTTAFold predicts structures solely through computational methods, sometimes eliminating the need for experimental measurements. AlphaFold 2, released in 2021, marked a significant breakthrough for deep learning in biology by unveiling a vast array of previously unknown protein structures. It quickly became a valuable tool for researchers working to understand everything from cellular structures [166] to tuberculosis [167]. It also inspired the development of other biological deep learning tools. Most notably, the biochemist David Baker and his team at the University of Washington developed a competing algorithm in 2021 called RoseTTAFold, which, like AlphaFold2, predicts protein structures from sequence data. Both systems have since been enhanced with new features. RoseTTAFold Diffusion is designed to create new proteins that do not exist in nature, while AlphaFold Multimer focuses on the interaction of multiple proteins. These advancements have propelled the development of numerous complementary tools that contextualize biochemical data, screen for protein interactions, and aid in experimental structure elucidation. Furthermore, the predictions from these tools have been integrated into publicly accessible databases, fostering widespread access and collaboration.

Proteins, intricate molecular machines honed by evolution, are built from a repertoire of 20 canonical amino acids, intricately arranged to yield diverse structures crucial for biological functions. Understanding a protein's 3D structure is paramount, as it dictates its functional properties; for instance, an enzyme's precise folding enables effective catalysis. Thus, deciphering protein structures not only determines their biological roles but also sheds light on disease-related mutations and their impacts. A longstanding aspiration in structural biology has been the computational prediction of protein structures, circumventing the laborious and expensive experimental methods. Milestones such as the Critical Assessment of Structure Prediction (CASP) [168] have gauged progress in this domain. AlphaFold's breakthrough at the 13th CASP competition, and subsequent advancements like AlphaFold2 and RoseTTAFold at 14th CASP competition, harnessed the pattern recognition prowess of machine-learning algorithms, trained on vast structural data repositories like the Protein Data Bank (PDB) [169]. These algorithms, unencumbered by prior exposure to certain proteins, demonstrated remarkable accuracy in structure prediction.

Following the 14th CASP competition, a proliferation of AI-enabled structure predictors has emerged. These predictors employ diverse strategies but share a common goal of understanding spatial proximity among amino acids by tracing evolutionary relationships. Multiple sequence alignment structure predictors (MSA-SPs), exemplified by AlphaFold 2 and RoseTTAFold, analyze co-evolutionary signals gleaned from input sequences to predict structures. In contrast, protein language model structure predictors (pLM-SPs), exemplified by ESMFold [170] and OmegaFold [171], embed evolutionary insights directly into their algorithms, eliminating the need for explicit MSA generation.

AlphaFold 3 [172], a successor to previous AlphaFold models, was released in 2024 by Google DeepMind. This new version extends its capabilities by predicting the structures of nearly all biological molecules and modeling their interactions. While researchers have previously developed specialized computational methods for modeling interactions between specific types of biological molecules, AlphaFold 3 is the first system capable of predicting interactions between almost all molecular types with state-of-the-art performance. The properties and functions of molecules in biological systems typically depend on their interactions with other molecules. Experimental methods to understand these interactions can take years and be prohibitively expensive. However, if these interactions can be accurately estimated computationally, biological research can be significantly accelerated. For instance, researchers looking for a promising drug candidate that binds a specific protein site can use computational systems like AlphaFold 3 to test potential drug molecules efficiently.

Other subcategories of BDTs include:

## Table 10. Other subcategories of BDTs

| Category | Description | Examples |
|---|---|---|
| Protein sequence design tools | Also known as 'inverse folding tools,' predict the sequence of a protein with a user-specified structure and/or functional property, such as binding to a target. These tools play a crucial role in designing proteins tailored to specific requirements. | Rosetta, RoseTTAFold, RF Diffusion |
| Small molecule design tools | Designed to predict molecular structures with specific profiles, such as generating drugs that provoke desired biological responses while maintaining acceptable pharmacokinetic properties. These tools are essential in drug discovery and development processes. | REINVENT 4, Chemistry42 |

| Category | Description | Examples |
| --- | --- | --- |
| Vaccine design tools | Pivotal in predicting protective antigens or vaccine subunits from the protein or proteome of target pathogens. By identifying vaccine candidates, these tools contribute significantly to the development of effective vaccines against infectious diseases. | LinearDesign, VSeq-Toolkit |
| Viral vector design tools | Focus on predicting the amino acid sequences of virus capsids with the aim of optimizing them as delivery vectors. These vectors are crucial in gene therapy and vaccine development, enabling the efficient delivery of therapeutic genes or vaccine antigens into target cells. | VSeq-Toolkit |
| Genetic modification tools | Analyze genetic sequences to identify sequence features or optimize them for specific purposes. These tools aid in genetic engineering applications by facilitating the modification of DNA sequences to achieve desired outcomes. | OpenCRISPR-1, ZFDesign |
| Genome assembly tools | Play a vital role in assembling genomes from multiple short reads generated by DNA sequencing technologies. These tools contribute to genome sequencing projects by reconstructing complete genome sequences from fragmented data. | DeepConsensus |
| Toxicity prediction/detection tools | Designed to predict or detect the molecular toxicity of given molecules or metabolites. These tools are valuable in drug safety assessment and environmental toxicology, aiding in the identification of potentially harmful substances. | TOXSCAPE, GENESCAPE |
| Pathogen property prediction tools | Predict or detect features of pathogens, such as propensity for zoonotic spillover or virulence. These tools are crucial in infectious disease surveillance and control, providing insights into the behavior and potential risks associated with pathogens. | MP4 |
| Host-pathogen interaction prediction tools | Focus on predicting protein-protein interactions between hosts and pathogenic agents. By elucidating the mechanisms of host-pathogen interactions, these tools contribute to understanding disease pathogenesis and identifying potential therapeutic targets. | HPIPred, deepHPI |

| Category | Description | Examples |
|---|---|---|
| Immunological system modeling tools | Replicate components of the human immune system to predict immune responses, such as T-cell receptor epitope recognition. These tools aid in vaccine design and immunotherapy development by simulating immune responses to pathogens or therapeutic agents. | SIMMUNE |
| Experimental design/planning tools | Generate designs for experiments based on predefined objectives, optimizing experimental variables and methods to achieve desired outcomes. These tools streamline the experimental process, improving efficiency and data quality. | The Experimental Design Assistant (EDA) |
| Experimental simulation tools | Simulate and predict experimental outcomes in silico, aiding in the design and interpretation of experiments. By providing insights into potential experimental outcomes, these tools inform experimental planning and hypothesis testing. | PhET, BioSimulators |
| Autonomous experimental platforms | Conduct experiments without human intervention, utilizing laboratory automation equipment to perform physical tests, modeling, or data mining. These platforms enhance experimental throughput and reproducibility, accelerating scientific research and discovery. | BO algorithm with expected Improvement based (EI-based) policy |

# Risks, Limitations and Future Directions

While recent advancements in AI have enabled rapid progress in the life sciences, it also has several limitations and presents potential risks.

## Risks and Limitations

### Inaccurate outputs from AI models

The effectiveness of AI tools relies heavily on the quality of their algorithms and the data they are trained on. When these algorithms contain errors or the datasets are biased or incomplete, the AI models can produce inaccurate outputs. If the models and logic underlying an AI algorithm are incorrect, the AI's predictions or recommendations will also be incorrect. This can occur due to coding errors, incorrect assumptions in the model design, or inadequate tuning of the model parameters. AI models learn from the data they are trained on. If the training data is biased (e.g., over-represents certain conditions or populations) or incomplete (e.g., missing critical variables or having insufficient diversity), the model's outputs will reflect these shortcomings. This means the AI could give incorrect advice or predictions, which in biological experiments can lead to wasted time and resources as researchers follow flawed directions. The inaccuracies can misguide researchers, causing them to conduct experiments based on false premises. This not only wastes valuable resources like time, money, and materials but can also delay scientific progress.

## Development of harmful biological agents

AI models have the potential to assist in the creation and distribution of harmful biological agents. They could, for example, enable an actor to design a biological agent with favorable properties [173] and modify the agent's delivery mechanism in a manner that optimizes infectious doses and ensures environmental survival [174].This possibility raises significant biosecurity concerns. Amateur users are unlikely to utilize BDTs, but experts with malicious intent could leverage their scientific training and specific AI models to design new pathogens, develop synthetic DNA strands that evade screening measures, or enhance the efficiency of bioweapon production [175]. As with any AI system, BDTs depend on the quality of their training data, which can sometimes be limited by incompleteness or unintentional biases. While BDTs have been used to digitally generate potentially risky genetic sequences, research has yet to show if the synthesized sequences could be used to create harmful biological agents. Establishing empirical baselines metrics is essential for conducting risk assessments and tracking changes in risk over time. In AI applications within the life sciences, these metrics and baselines are not yet defined. To assess this risk, we need to systematically evaluate current AI systems' abilities to generate new sequences versus enhancing existing ones.

## Ethics in AI for Life Sciences

The integration of artificial intelligence (AI) in life sciences presents significant ethical challenges that must be addressed to ensure responsible and beneficial use. Key ethical considerations include data privacy, informed consent, and the potential for bias in AI algorithms. Ensuring data privacy is paramount, as AI systems often require access to vast amounts of sensitive biological and medical data. This necessitates robust data protection measures and compliance with legal standards to prevent misuse and unauthorized access [176]. Informed consent is another critical issue, as individuals must be fully aware of how their data will be used and the potential implications of AI-driven analyses [177]. Additionally, AI algorithms can inadvertently perpetuate or exacerbate existing biases if the training data is not representative of diverse populations, leading to inequitable outcomes in healthcare and research [176]. Addressing these ethical concerns requires a multi-faceted approach, including rigorous testing of AI systems, transparency in AI operations, and the establishment of ethical guidelines and governance frameworks to guide the development and deployment of AI in life sciences [178]. By prioritizing these ethical considerations, we can harness the transformative potential of AI while safeguarding human rights and promoting equitable access to its benefits.

# Future Directions

## Introduction of new benchmarks

Recent studies have highlighted the shortcomings of existing benchmarks in evaluating LLMs for clinical applications [179,180]. Traditional benchmarks, which focus mainly on accuracy in medical question-answering, fail to capture the full range of clinical skills necessary for LLMs [181]. Critics argue that using human-centric standardized medical exams to evaluate LLMs is insufficient, as passing these tests does not reflect the nuanced expertise required in real-world clinical settings [181].

There is a growing consensus on the need for more comprehensive benchmarks. These new benchmarks should assess capabilities such as sourcing information from authoritative medical references, adapting to the evolving medical knowledge landscape, and clearly communicating uncertainties [181,182]. To further enhance their relevance, benchmarks should include scenarios that test an LLM's performance in real-world applications and its ability to adapt to feedback from clinicians while maintaining robustness. Given the sensitive nature of healthcare, these benchmarks should evaluate factors like fairness, ethics, and equity, which are crucial yet challenging to quantify [181]. By expanding benchmarks to encompass scientific domains, especially the biological domain, we can ensure that LLMs are rigorously evaluated across a broad spectrum of applications, thereby promoting their responsible and effective use in advancing scientific and medical knowledge.

## Red, blue and violet teaming

Due to increasing concerns about the safety, security, and trustworthiness of Generative AI models, both practitioners and regulators emphasize the importance of AI red-teaming [183]. Originally from cybersecurity, red-teaming involves adopting an adversary's perspective to find vulnerabilities. In AI, this means simulating attacks on AI applications to identify weaknesses and develop preventive measures [184]. For example, red teams can simulate backdoor attacks or data poisoning to test the AI model's defenses. Prompt injection, a common attack on generative AI models like LLMs, tricks the model into producing harmful content. Red teams can also prompt AI systems to extract sensitive information from training data.

Blue teaming, which focuses on defending against these simulated attacks, and purple teaming, which combines both red and blue teams for a comprehensive security assessment [185]. However, as AI systems continuously evolve, these strategies might be insufficient, especially in critical sectors like the life sciences [186].

Violet teaming goes further by pairing red and blue teams to build resilient systems that intend to simultaneously minimize harm and maximize benefit using the very technology that poses potential security risks [187]. In the life sciences, this might involve using AI models to screen for harmful sequences generated by the models themselves, preventing them from being produced and shared with the end user.

Additionally, Machine Learning Security Operations (MLSecOps) could play a crucial role in ensuring the safety of AI models in the life sciences by employing machine learning (ML) techniques to protect against cyber threats and secure AI/ML models [188]. MLSecOps focuses on encrypting sensitive genome data, detecting ransomware and Trojan attacks, and ensuring the integrity of ML algorithms used in critical applications. It also addresses vulnerabilities in software and IoT devices within biotechnology labs, enhances supply chain security, and mitigates biases in healthcare ML systems.

# Conclusion

The integration of computing technologies into the life sciences has profoundly transformed the field, enabling unprecedented advancements in biological research and applications. From the early days of population genetics calculations in the 1950s to the sophisticated AI-driven models of today, the evolution of computational tools has paralleled and propelled the growth of life sciences.

## Historical Milestones and Technological Advancements

The journey began with the use of primitive computers for biological modeling, such as Alan Turing's work on morphogenesis and the MANIAC computer's genetic code measurements. The 1960s and 1970s saw the rise of computational biology, driven by protein crystallography and the development of bioinformatics software like COMPROTEIN. The advent of dynamic programming algorithms for sequence alignment and the shift from protein to DNA analysis marked significant milestones.

## The Genomic Era and Beyond

The 1980s and 1990s were pivotal, with the development of gene targeting techniques, the polymerase chain reaction (PCR), and the emergence of bioinformatics software suites. The completion of the Haemophilus influenzae genome and the human genome project ushered in the genomic era, leading to the creation of specialized software for whole-genome sequencing.

## Artificial Intelligence and Machine Learning

The recent decades have witnessed the integration of artificial intelligence (AI) and machine learning (ML) into life sciences, revolutionizing data analysis, drug discovery, and personalized medicine. AI models, from expert systems like DENDRAL and MYCIN to modern deep learning architectures, have enhanced our ability to predict protein structures, analyze genomic data, and design novel biological entities.

## Emerging Technologies and Future Directions

Emerging technologies such as cloud computing, big data analytics, and the Internet of Things (IoT) are further enhancing the capabilities of life sciences research. Cloud-based high-performance computing enables complex data analysis and reduces research cycles, while IoT facilitates real-time data collection and patient monitoring.

## Challenges and Ethical Considerations

Despite these advancements, challenges remain. The accuracy of AI models depends on the quality of training data, and there are significant ethical concerns regarding data privacy and the potential misuse of AI in creating harmful biological agents. Addressing these challenges requires robust ethical frameworks, continuous monitoring, and the development of explainable AI systems.

Altogether, the integration of computing in the life sciences has not only accelerated research but also opened new frontiers in understanding and manipulating biological systems. As we move forward, the synergy between computational technologies and life sciences will continue to drive innovation, offering new solutions to complex biological problems and improving human health. The future promises even greater advancements as we harness the power of AI, cloud computing, and other emerging technologies to explore the intricacies of life at unprecedented scales.

# Glossary

- **Algorithm:** A step-by-step procedure or formula for solving a problem, often used in computer programming and computational biology.

- **AlphaFold:** An AI program developed by DeepMind that predicts protein structures with high accuracy.

- **Autoencoders:** A type of artificial neural network used to learn efficient codings of unlabeled data.

- **Bioinformatics:** The application of computer technology to the management and analysis of biological data.

- **Computational Biology:** A field that uses mathematical models, algorithms, and computational techniques to understand and analyze biological systems.

- **CRISPR:** A technology used for editing genomes, allowing researchers to alter DNA sequences and modify gene function.

- **Deep Learning:** A subset of machine learning involving neural networks with many layers.

- **DNA Sequencing:** The process of determining the nucleic acid sequence – the order of nucleotides in DNA.

- **Fourier Analysis:** A mathematical technique used to transform signals between time (or spatial) domain and frequency domain, applied in protein crystallography to determine structures.

- **Fixed-size vector representation:** A numerical representation of a fixed length that encapsulates the information or features extracted from a variable-length input. In machine learning and natural language processing (NLP), fixed-size vector representations are commonly used to represent textual or sequential data.

- **Genome:** The complete set of genes or genetic material present in a cell or organism.

- **Genomics:** The study of genomes, the complete set of DNA within an organism, including its structure, function, evolution, and mapping.

- **Generative Adversarial Networks (GANs):** A class of machine learning systems where two neural networks contest with each other in a game.

- **Machine Learning (ML):** A subset of artificial intelligence (AI) that involves the development of algorithms that allow computers to learn from and make predictions based on data.

- **Maximum Likelihood Methods:** Statistical methods for estimating the parameters of a model, used in phylogenetic inference to determine the most likely tree structure.

- **Metagenomics:** The study of genetic material recovered directly from environmental samples.

- **Multiple Sequence Alignment (MSA):** A method used to align three or more biological sequences to identify regions of similarity that may indicate functional, structural, or evolutionary relationships.

- **Natural Language Processing (NLP):** The ability of a computer program to understand human language as it is spoken.

- **Needleman-Wunsch Algorithm:** An algorithm used for pairwise sequence alignment that employs dynamic programming to find the optimal alignment between two sequences.

- **Neural Networks:** A series of algorithms that mimic the operations of a human brain to recognize relationships between vast amounts of data.

- **Next-Generation Sequencing (NGS):** High-throughput sequencing technologies that allow for rapid sequencing of DNA or RNA samples.

- **PCR (Polymerase Chain Reaction):** A method widely used in molecular biology to make several copies of a specific DNA segment.

- **Phylogenetics:** The study of the evolutionary history and relationships among individuals or groups of organisms.

- **Proteomics:** The large-scale study of proteins, particularly their structures and functions.

- **Reinforcement Learning:** A type of machine learning where an agent learns to behave in an environment by performing actions and seeing the results.

- **Systems Biology:** An approach in biomedical research to understanding the larger picture by putting its pieces together (holism instead of reductionism).

- **Transcriptomics:** The study of the complete set of RNA transcripts produced by the genome.

- **Unsupervised Learning:** A type of machine learning that looks for previously undetected patterns in a data set with no pre-existing labels.

123. **Bilingual Language Model for Protein Sequence and Structure**
Michael Heinzinger, Konstantin Weissenow, Joaquin Gomez Sanchez, Adrian Henkel, Milot Mirdita, Martin Steinegger, Burkhard Rost
*bioRxiv* (2024-03-24) https://www.biorxiv.org/content/10.1101/2023.07.23.550085v2
DOI: 10.1101/2023.07.23.550085

124. **Generative Antibody Design for Complementary Chain Pairing Sequences through Encoder-Decoder Language Model**
Simon KS Chu, Kathy Y Wei
*arXiv.org* (2023-01-06) https://arxiv.org/abs/2301.02748v4

125. **Fold2Seq: A Joint Sequence(1D)-Fold(3D) Embedding-based Generative Model for Protein Design**
Yue Cao, Payel Das, Vijil Chenthamarakshan, Pin-Yu Chen, Igor Melnyk, Yang Shen
*arXiv* (2021-06-24) http://arxiv.org/abs/2106.13058
DOI: 10.48550/arxiv.2106.13058

126. **Predicting Retrosynthetic Reaction using Self-Corrected Transformer Neural Networks**
Shuangjia Zheng, Jiahua Rao, Zhongyue Zhang, Jun Xu, Yuedong Yang
*arXiv* (2019-07-02) http://arxiv.org/abs/1907.01356
DOI: 10.48550/arxiv.1907.01356

127. **DNABERT: pre-trained Bidirectional Encoder Representations from Transformers model for DNA-language in genome**
Ji Y, Zhou Z, Liu H, Davuluri Rv
*Bioinformatics (Oxford, England)* (2021-08-09) https://pubmed.ncbi.nlm.nih.gov/33538820/
DOI: 10.1093/bioinformatics/btab083

128. **DNABERT-2: Efficient Foundation Model and Benchmark For Multi-Species Genome**
Zhihan Zhou, Yanrong Ji, Weijian Li, Pratik Dutta, Ramana Davuluri, Han Liu
*arXiv* (2024-03-18) http://arxiv.org/abs/2306.15006
DOI: 10.48550/arxiv.2306.15006

129. **iEnhancer-BERT: A Novel Transfer Learning Architecture Based on DNA-Language Model for Identifying Enhancers and Their Strength**
springerprofessional.de
https://www.springerprofessional.de/en/ienhancer-bert-a-novel-transfer-learning-architecture-based-on-d/23365796

130. https://www.researchgate.net/publication/362540943_MoDNA_motif-oriented_pre-training_for_DNA_language_model

131. **GENA-LM: A Family of Open-Source Foundational Models for Long DNA Sequences**
Veniamin Fishman, Yuri Kuratov, Maxim Petrov, Aleksei Shmelev, Denis Shepelin, Nikolay Chekanov, Olga Kardymon, Mikhail Burtsev
*bioRxiv* (2023-06-13) https://www.biorxiv.org/content/10.1101/2023.06.12.544594v1
DOI: 10.1101/2023.06.12.544594

132. **The Nucleotide Transformer: Building and Evaluating Robust Foundation Models for Human Genomics**
Hugo Dalla-Torre, Liam Gonzalez, Javier Mendoza Revilla, Nicolas Lopez Carranza, Adam Henryk Grzywaczewski, Francesco Oteri, Christian Dallago, Evan Trop, Hassan Sirelkhatim, Guillaume Richard, … Thomas Pierrot
*bioRxiv* (2023-01-15) https://www.biorxiv.org/content/10.1101/2023.01.11.523679v1
DOI: 10.1101/2023.01.11.523679

(2023-10-10) https://www.nist.gov/artificial-intelligence/executive-order-safe-secure-and-trustworthy-artificial-intelligence

184. **What Does AI Red-Teaming Actually Mean?**
Tessa Baker
*Center for Security and Emerging Technology* (2023-10-24)
https://cset.georgetown.edu/article/what-does-ai-red-teaming-actually-mean/

185. **Purple Teaming: A comprehensive and collaborative approach to cyber security**
Erik Van Buggenhout
*Cyber Security: A Peer-Reviewed Journal* (2024)
https://ideas.repec.org//a/aza/csj000/y2024v7i3p207-216.html

186. https://www.researchgate.net/publication/372592054_Violet_Teaming_AI_in_the_Life_Sciences_A_Preprint

187. **The Promise and Peril of Artificial Intelligence -- Violet Teaming Offers a Balanced Path Forward**
Alexander J Titus, Adam H Russell
*arXiv* (2023-08-27) http://arxiv.org/abs/2308.14253
DOI: 10.48550/arxiv.2308.14253

188. **Integrating MLSecOps in the Biotechnology Industry 5.0**
Naseela Pervez, Alexander J Titus
*IntechOpen* (2024-05-10) https://www.intechopen.com/online-first/89417
ISBN: 9780850144840